\documentclass{bmcart}

\usepackage{amsthm,amsmath}
\usepackage[utf8]{inputenc} 
\usepackage{tabularx}
\usepackage{csquotes}
\usepackage[normalem]{ulem}
\usepackage{booktabs}
\usepackage{cancel}
\usepackage{xr}
\usepackage{url}
\makeatletter
\newcommand*{\addFileDependency}[1]{
  \typeout{(#1)}
  \@addtofilelist{#1}
  \IfFileExists{#1}{}{\typeout{No file #1.}}
}
\makeatother

\newcommand*{\myexternaldocument}[1]{%
    \externaldocument{#1}%
    \addFileDependency{#1.tex}%
    \addFileDependency{#1.aux}%
}

\myexternaldocument{supplementary_material}

\usepackage{graphicx}

\startlocaldefs
\endlocaldefs

\begin{document}

\begin{frontmatter}

\begin{fmbox}
\dochead{Research}


\title{Venture capital investments through the lens of network and functional data analysis}


\author[
  addressref={aff1,aff6}, 
  noteref={n1},
  email={XXX}
]{\inits{C.E.}\fnm{Christian} \snm{Esposito}}
\author[
  addressref={aff3},
  noteref={n1},
  email={marco.gortan@studbocconi.it}
]{\inits{M.G.}\fnm{Marco} \snm{Gortan}}
\author[
  addressref={aff1,aff6},
  corref={aff1,aff6},   
  noteref={n1},
  email={l.testa@sssup.it}
]{\inits{L.T.}\fnm{Lorenzo} \snm{Testa}}
\author[
  addressref={aff1,aff2},                   
  corref={aff1,aff2},                       
  email={fxc11@psu.edu}   
]{\inits{F.C.}\fnm{Francesca} \snm{Chiaromonte}}
\author[
  addressref={aff1}, 
  email={g.fagiolo@santannapisa.it}
]{\inits{G.F.}\fnm{Giorgio} \snm{Fagiolo}}
\author[
  addressref={aff1,aff4},
  email={andrea.mina@santannapisa.it}
]{\inits{A.M.}\fnm{Andrea} \snm{Mina}}
\author[
  addressref={aff5},
  email={giulio.rossetti@isti.cnr.it}
]{\inits{G.R.}\fnm{Giulio} \snm{Rossetti}}


\address[id=aff1]{
  \orgdiv{Institute of Economics \& EMbeDS},             
  \orgname{Sant'Anna School of Advanced Studies},          
  \city{Pisa},                              
  \cny{IT}                                    
}
\address[id=aff6]{
  \orgdiv{Department of Computer Science},             
  \orgname{University of Pisa},          
  \city{Pisa},                              
  \cny{IT}                                    
}
\address[id=aff3]{
  \orgname{Bocconi University},          
  \city{Milan},                              
  \cny{IT}                                    
}
\address[id=aff2]{%
  \orgdiv{Department of Statistics \& Huck Institutes of the Life Sciences},
  \orgname{Penn State University},
  \city{University Park},
  \cny{PA, USA}
}
\address[id=aff4]{
  \orgdiv{Centre for Business Research},             
  \orgname{University of Cambridge},          
  \city{Cambridge},                              
  \cny{UK}                                    
}
\address[id=aff5]{
  \orgdiv{KDD Lab.},             
  \orgname{ISTI-CNR},          
  \city{Pisa},                              
  \cny{IT}                                    
}


\begin{artnotes}
\note[id=n1]{Equal contributor} 
\end{artnotes}

\end{fmbox}


\begin{abstractbox}

\begin{abstract} 
In this paper we characterize the performance of venture capital-backed firms based on their ability to attract investment. The aim of the study is to identify relevant predictors of success built from the network structure of firms' and investors' relations. Focusing on deal-level data for the health sector, we first create a bipartite network among firms and investors, and then apply functional data analysis (FDA) to derive progressively more refined indicators of success captured by a binary, a scalar and a functional outcome. More specifically, we use different network centrality measures to capture the role of early investments for the success of the firm. Our results, which are robust to different specifications, suggest that success has a strong positive association with centrality measures of the firm and of its large investors, and a weaker but still detectable
association with centrality measures of small investors and features describing firms as knowledge bridges. Finally, based on our analyses, success is not associated with firms' and investors' spreading power (harmonic centrality), nor with the tightness of investors' community (clustering coefficient) and spreading ability (VoteRank).
\end{abstract}




\end{abstractbox}
%

\end{frontmatter}



\section*{Introduction}
In their pursuit of entrepreneurial opportunities, new firms rely on a variety of financing sources. When internal means of financing (e.g. owner capital and cash flow) are insufficient to support the growth of the business, firms will seek external capital \cite{hall2010financing,cosh2009outside,mina2013demand}. In the context of high-risk and technology-intensive sectors, venture capital (VC) is the kind of external capital that can provide not only financial resources, but also substantial knowledge of product markets, and useful connections with other firms and investors \cite{da2013survey}. 

In the VC investment model, success implies a positive exit outcome of the investee firm through an IPO or trade sale, generating (ideally optimal) returns for the investors \cite{gompers1995optimal,poulsen2008moving}. From the viewpoint of the firm, there are indeed other indicators of positive performance, which include new patents \cite{lerner2000assessing,lahr2016venture}, new products \cite{hellmann2000interaction}, and growth by number of employees and sales \cite{bertoni2011venture}, depending on the stage of development of the business. 

In this study we take a different approach, and consider the ability of the firm to raise funds as an indication that the firm is achieving those milestones that mark the path to exit events, and remains through time a promising vehicle of future returns. We implement this procedural view of success by investigating specific features of the network of firms and investors built from deal-level data. 

A number of studies employ network tools to describe interactions among these economic agents. For instance, Bonaventura \textit{et al.} \cite{bonaventura2020predicting} use networks measures to link the likelihood of long-term positive economic performance to the flow of employees (and the associated transfer of know-how) across firms. Network tools can also be used to assess the probability that an investor will invest in a certain firm \cite{liang2016predicting}, and to study how suppliers of finance can exchange opportunities to invest in a portfolio firm, spread financial risk and share knowledge \cite{bygrave1988structure}. Investors' networks appear to be especially important for the likelihood of positive returns and successful exits \cite{hochberg2007whom}. In this contribution we use network analysis to build, first, a binary definition of success by which we partition funding trajectories in two clusters capturing a `high' and a `low' funding regime. Then, we derive richer characterizations of firm success by considering the total amount of money raised as a scalar outcome, and the funding trajectory itself as a functional outcome.

We retrieve data from CB Insights \cite{cbinsights}, which provides records of transactions in venture capital markets and stock market listings from 1949. Notably, the very first transaction in our dataset corresponds to the public listing of five Japanese firms in May 1949, while the first recorded venture capital transaction concerns an investment by Greylock Partners in Worthington Biochemical -- labelled by CB Insights as `Growth Equity' in 1967. However, since data until 2000 are incomplete, we focus on the period 2000–2020 in order to minimize the impact of missing data on our analysis. Moreover, as different sectors may follow different investment patterns \cite{dushnitsky2006does}, we restrict our analysis to the healthcare sector only, which is one of the richest in venture capital investments and one that has shown to be among the least sensitive to market oscillations \cite{pisano2006science}.

The reminder of this article is organized as follows. After a description of the complex, time-varying structure of the bipartite network of investors and firms, we show how our definitions of success can be related to standard definitions in the literature. Next, we introduce statistics computed on the projections of the bipartite network, and show their association with our definitions of success. We do this using a temporal window of 10 years along which we consider a firm's financing rounds after its first investment, and demonstrate the advantages of our approach. Since the length of the temporal window employed in our analysis is somewhat arbitrary, we repeat it with lengths varying from 5 to 12 years -- ascertaining the robustness of our findings. We apply a similar stability check varying the set of covariates employed in our regression models. Finally, we discuss our main results and provide some concluding remarks.

\section*{Network characterization}
We build a \textit{bipartite network} with 83258 nodes divided into investors (32796) and firms (50462) -- notably, the two categories are not exclusive; 2155 nodes are labeled as both firms and investors. An investor and a firm node are connected by an undirected link whenever there exists a transaction in the CB Insights' database regarding the two, i.e.~a record of the investor investing in the firm, and for which we have information regarding the amount of money invested and the date of the deal. We end up with 63035 links. Because an investor can invest in the same firm multiple times, the network is also a \textit{multigraph}. Moreover, we account for the \textit{temporal} dimension in a cumulative fashion: once created, links persist, so
the snapshots of the graph that we compute for each year between 2000 and 2020 contain all the links (and their nodes) concerning transactions recorded prior to and including that year. 

By projecting the bipartite network on the firms' layer, we produce a projected graph which is employed to compute the statistics described in Table \ref{tab:network_statistics}. In particular, we assume that two firms are linked if they are invested in by the same investor within seven years. This time span balances a trade-off between the average VC investment window, which recent evidence places below five years \cite{sethuram2021multiple}, and the realistic advantages that may derive from the expertise of a common investor -- even after its exit from one of the firms. As a robustness check, we run our pipelines also linking two firms if they are invested by the same investor within either five or ten years (see supplementary material).

By projecting the bipartite network on the investors' layer, we produce a second projected graph. Here two investors are linked if they have invested in the same firm in the same financing round. In order to use also investors' centrality measures as potential predictors for firms' success, we compute the maximum, the minimum and the median of the centrality distribution of the `early' investors in each firm, i.e.~those who participated in the firm's first recorded funding round. 

\begin{table}
\caption{Statistics computed on the projected graphs of investors and firms. To link investors' centrality measures to firms, we consider investors involved in the first investment round of a given firm, and summarize the distributions of their statistics through maximum, minimum and median.}
  \begin{tabularx}{\textwidth}{lXl}
    \toprule
    \textbf{Covariate} & \textbf{Network interpretation} & \textbf{Note} \\ 
    \midrule
    Average neighbor degree & Affinity between neighbor nodes & \\
    Betweenness centrality & Role within flow of information & \\
    Closeness centrality & Spreading power (shortest average
    distance from all other nodes) & \\
    Clustering coefficient & Tight community & \\
    Core number & Importance within cluster & Computed only for firms\\
    Degree centrality & Influence & \\
    Eigenvector centrality & Influence & \\
    Harmonic centrality \cite{marchiori2000harmony}& Spreading power & \\
    Newman betweenness centrality \cite{newman2005measure} & Role within flow of information & \\
    Number of investors &  & Computed only for firms\\
    PageRank \cite{page1999pagerank} & Influence & \\
    VoteRank \cite{zhang2016identifying} & Best spreading ability & \\
    \bottomrule
  \end{tabularx}
\label{tab:network_statistics}
\end{table}

\begin{figure}
    \centering
    \includegraphics[width=0.96\textwidth]{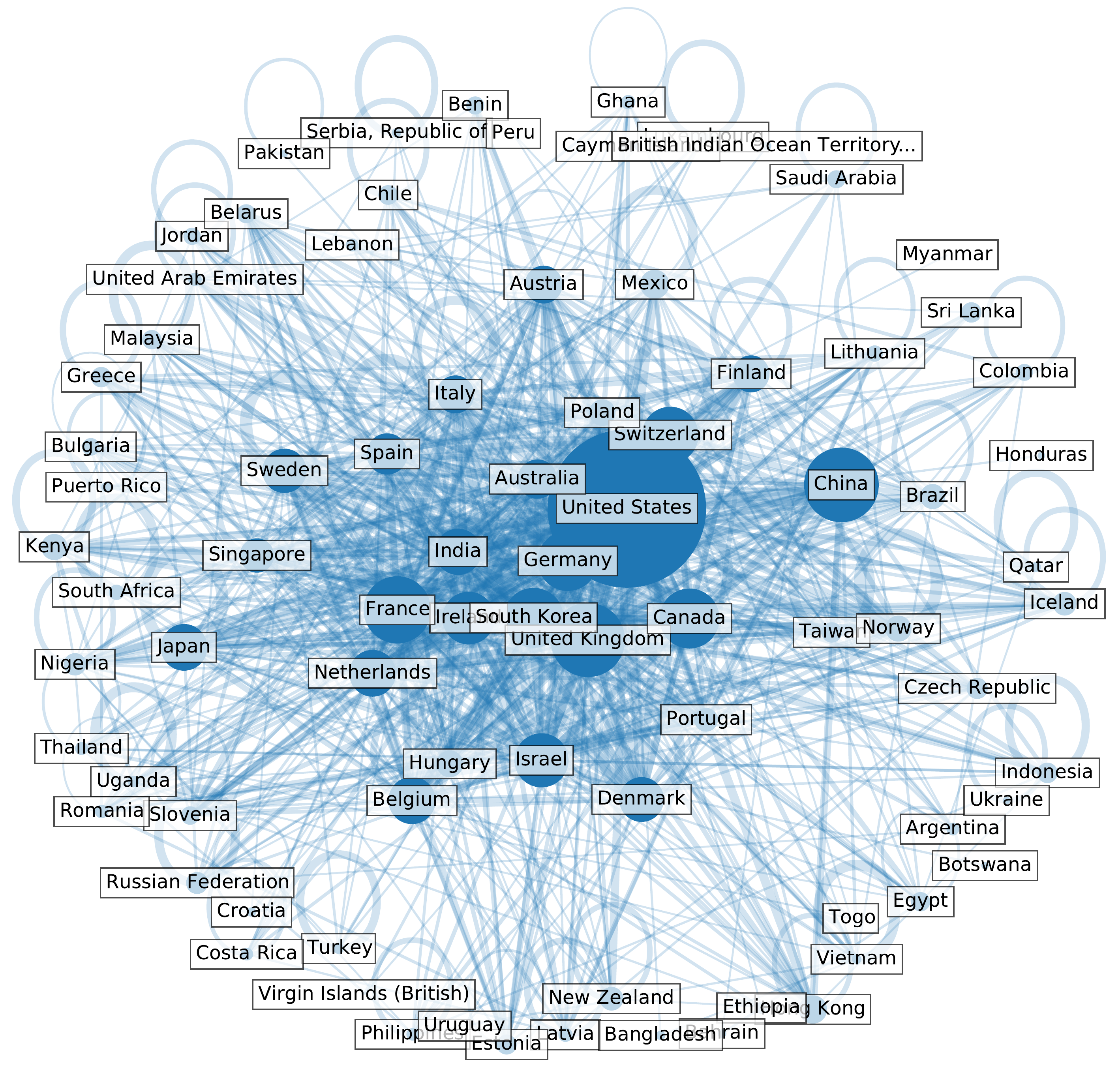}
  \caption{Projection of the bipartite network on the firms' layer, aggregated by country.}
  \label{fig:network}
\end{figure}

Figure \ref{fig:network} gives a snapshot of the firms' projected network in 2020, aggregated at the geographical resolution of countries. The size of each node, which represents a country, is given by the number of firms located in that country. 
Roughly 83\% of the firms are either North American or European (around 60\% belong to the US market), while the remaining 17\% is mostly composed of Asian firms. In terms of sub-sectoral specialization, around 50\% of the firms operate within the sub-sectors of Medical Devices \& Equipment, Biotechnology, and Internet Software \& Services – the Medical Devices \& Equipment sub-sector alone accounts for 20\% of the nodes in network. About 85\% of the firms in the network are either active or acquired, with the remaining portion being either inactive or having completed an IPO.

\section*{Results}
\subsection*{Assessing centrality measures against standard definitions of success}

We start by assessing whether the centrality measures computed from our networks correlate with firms' success according to a standard definition, reproducing the exercise presented in Bonaventura \textit{et al.} \cite{bonaventura2020predicting}.
We classify a firm as successful if, in a time window of given size from its first investment, (i) it has been acquired, (ii) it has been listed in the public market, or (iii) it has merged with another firm. Therefore, as compared to Bonavenura \textit{et al.} \cite{bonaventura2020predicting}, we differ both in the creation of the networks and, slightly, in the definition of success (we do not label a firm as successful if it has made an acquisition).

Bonaventura {\em et al.} \cite{bonaventura2020predicting} start by considering the pool of \textit{open deals} available to investors each month -- these are the firms which (i) have not yet received funding, (ii) have not yet been acquired, and (iii) have not yet been listed in the stock exchange market. Next, they propose an investment strategy by picking, each month, the top $n$ firms in terms of closeness centrality, and check whether -- over a time span of 6, 7 or 8 years -- this strategy is more successful than just randomly picking $n$ firms from the pool of open deals. To assess statistical significance, the authors apply a test based on a hypergeometric null distribution (the number of successful firms in a random sample of size $n$ from a population whose overall size and prevalence of successful firms are known).

\begin{figure}
    \includegraphics[width=0.98\textwidth]{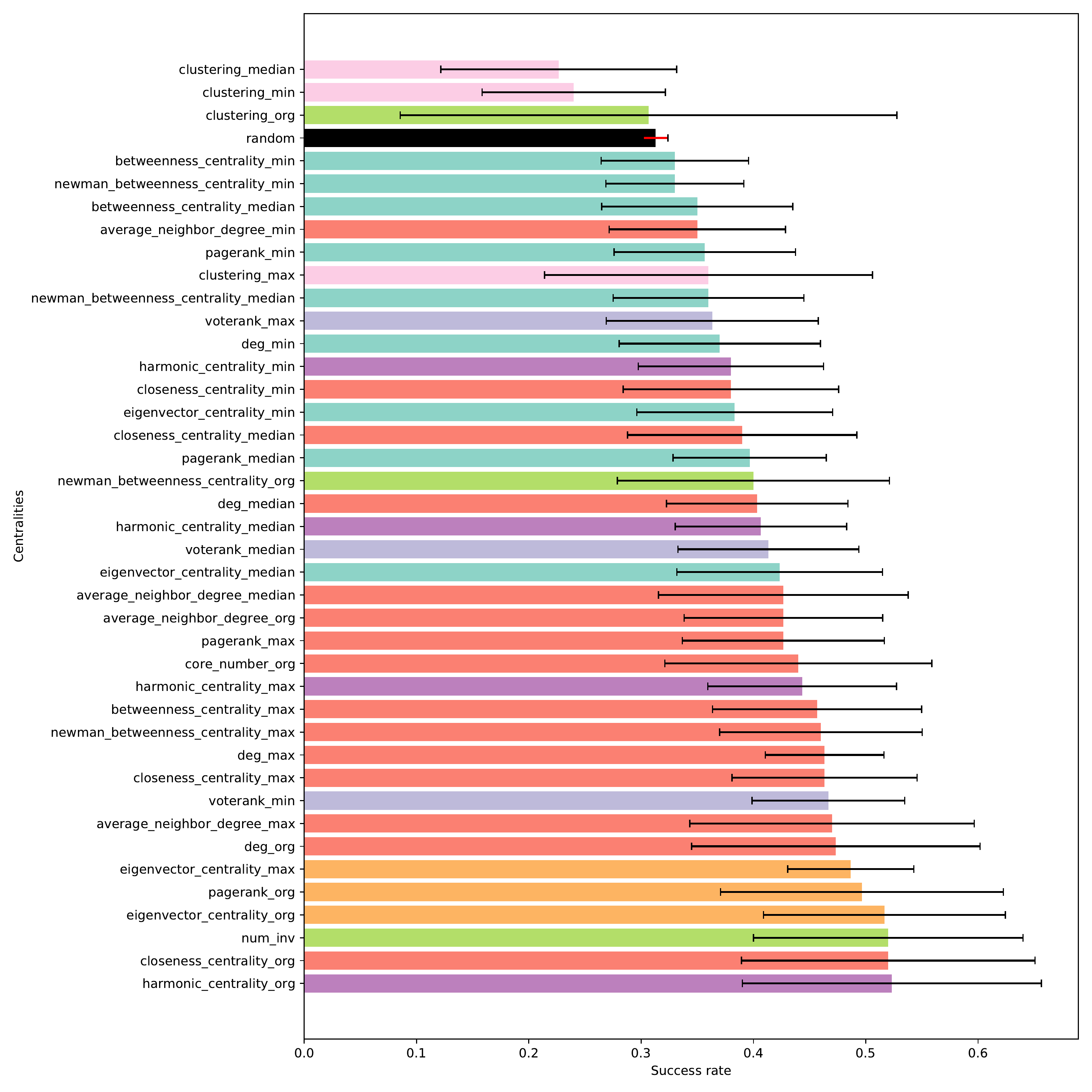}
  \caption{Success rate for different centrality measures adopting the methodology proposed by Bonaventura \textit{et al.} \cite{bonaventura2020predicting}. Bars represent mean success rates over $11$ years (from 2000 to 2010), and are color-coded according to the partition of centrality measures obtained from the dendrogram in Figure~\ref{fig:dendrogram}. Black lines mark $\pm1$ standard deviation intervals about the means.}
  \label{fig:bonav}
\end{figure}

In our analysis, we link firms and investors according to observed investments. Therefore, we act as an investor who has already seen the first investment in the firms of interest. Figure~\ref{fig:bonav} shows success rates computed using this approach for different centrality measures with $n=25$. For each measure, firms are sorted in descending order to pick the top $25$ -- except for VoteRank, which captures an inverse centrality, so firms are sorted in ascending order. Results are aggregated computing mean success rates and their standard deviations over $11$ years (from 2000 to 2010), and shown grouping centrality measures based on their correlation structure (see below). As a preliminary to analyses to be presented later in the paper, we note that group 4 emerging from the dendrogram in Figure~\ref{fig:dendrogram} contains centrality measures that produce similar
and rather good success rates; centrality measures from groups 3 and 5 produce more heterogeneous success rates, and centrality measures from group 1 produce poorer success rates. 

\subsection*{Extracting centrality signals}
\begin{figure}
  \includegraphics[width=0.98\textwidth]{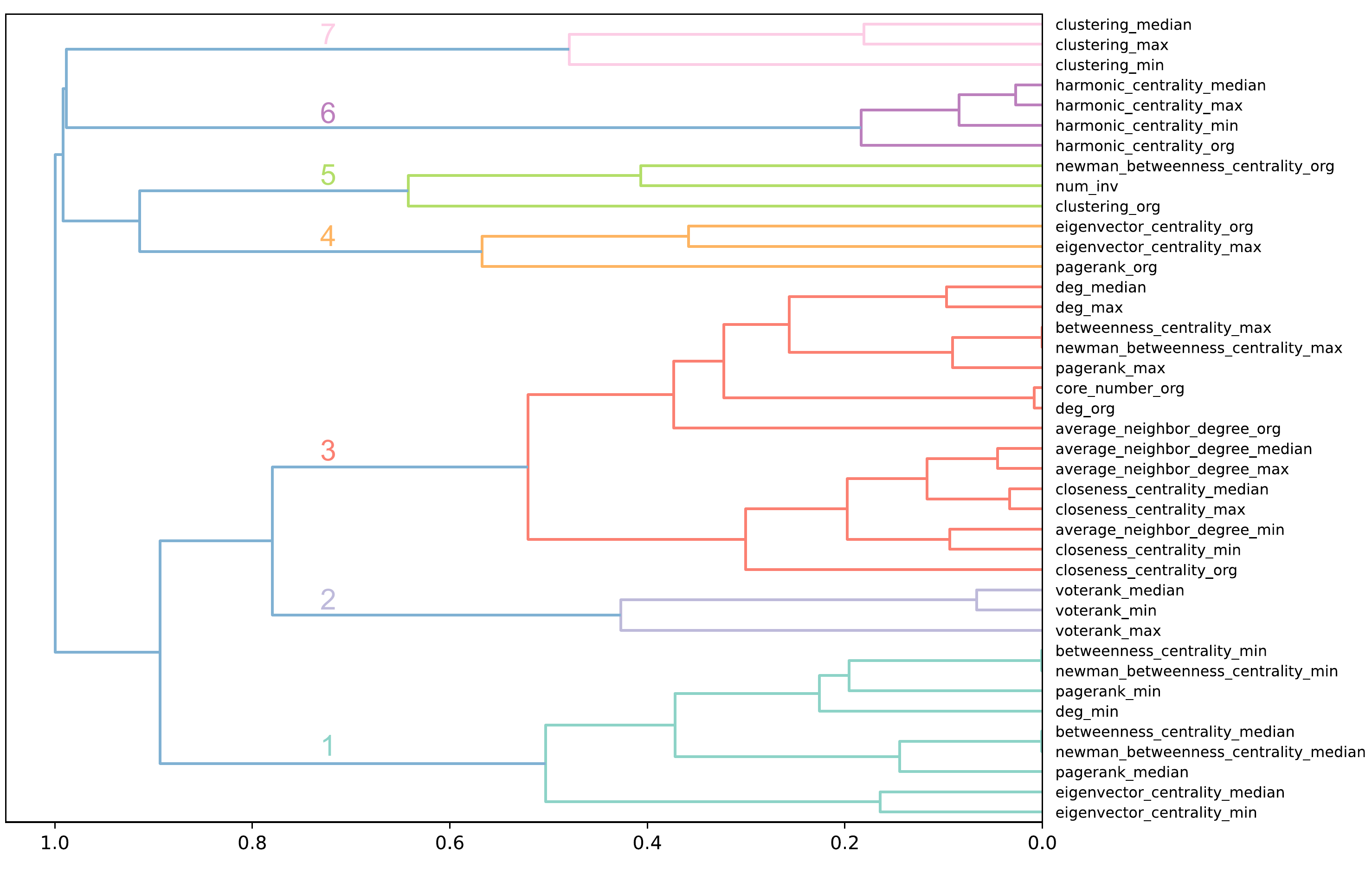}
  \caption{Dendrogram of firms' and investors' centrality measures (absolute correlation distance, complete linkage). Seven groups of features are highlighted. The first from the bottom contains centrality measures for smallest and median investors; the second includes investors' VoteRank statistics; the third contains firms' and investors' closeness and average neighborhood centrality, as well as measures of firms' and big investors' centrality; the fourth includes firms' and big investors' eigenvector centrality measures, as well as firms' PageRank; the fifth contains features describing firms as knowledge bridges; the sixth includes harmonic centrality measures; the seventh includes investors' clustering coefficients.}
  \label{fig:dendrogram}
\end{figure}

After assessing centrality measures against standard definitions of success, and before turning to regression exercises for alternative and progressively more refined success outcomes, we pre-process our covariates as follows. After log-transforming those that present markedly right-skewed distributions, we scale them all and analyze their correlation structure by building a feature dendrogram (Pearson absolute correlation, complete linkage; see Figure \ref{fig:dendrogram}). The dendrogram highlights seven groups. Within each of them our covariates are highly correlated, but across the groups we can identify distinct, fairly uncorrelated signals. The first group includes investors' minimum and median eigenvector, Newman betweenness, PageRank, betweenness centrality and investors' minimum degree. The second group comprises the maximum, the minimum and the median of investors' VoteRank. The third group, which is the largest one, contains information on firms' degree and core number and investors' summary statistics on average neighbor degree and closeness centrality, together with investors' maximum PageRank, Newman betweenness centrality, degree and betweenness centrality. The fourth group features firms' eigenvector and PageRank centrality and investors' maximum eigenvector centrality. The fifth group describes firms as knowledge bridges (number of investors, clustering coefficient and Newman betweenness centrality). The sixth group comprises firms' harmonic centrality and summary statistics of investors' harmonic centrality. Finally, the seventh group includes summary statistics of investors' clustering coefficients.

We leverage these groups to guide feature selection for our regression exercises. For each response type (binary, scalar, functional; more on that below), we reduce the initial set of covariates to seven predictors, selecting one per group through an exhaustive search for the combination that optimizes the goodness of fit. We later consider further (sub-optimal) combinations comprising one covariate per group, as a check on the stability of our analysis.

\subsection*{Refining the definition of success}

Each firm has its own funding history. After its birth, the firm collects resources over time, thus being characterized by the \textit{trajectory} of the amount of money it is able to attract. We treat these trajectories with tools from Functional Data Analysis (FDA), a field of statistics that studies observations that come in the form of functions over a continuous domain \cite{ramsay2005,kokoszka2017introduction}. Specifically, we focus on \textit{cumulative} functions of the money raised over time by each firm, which by construction are monotonically non-decreasing. These are \textit{aligned}, so that their time domain begins the year in which each firm receives its first investment (regardless of the calendar year it corresponds to). We restrict attention to firms that have received at least two investments during their time domain, as to avoid flat trajectories. Moreover, we only consider firms whose sub-sector is known. In the following, we show results obtained using a time window of 10 years from first investment, employing a total of 3072 firms. Subsequently, we also explore alternative window sizes (see Figure \ref{fig:n_firms}).

\subsubsection*{Two regimes}
\begin{figure}
    \includegraphics[width=0.60\textwidth]{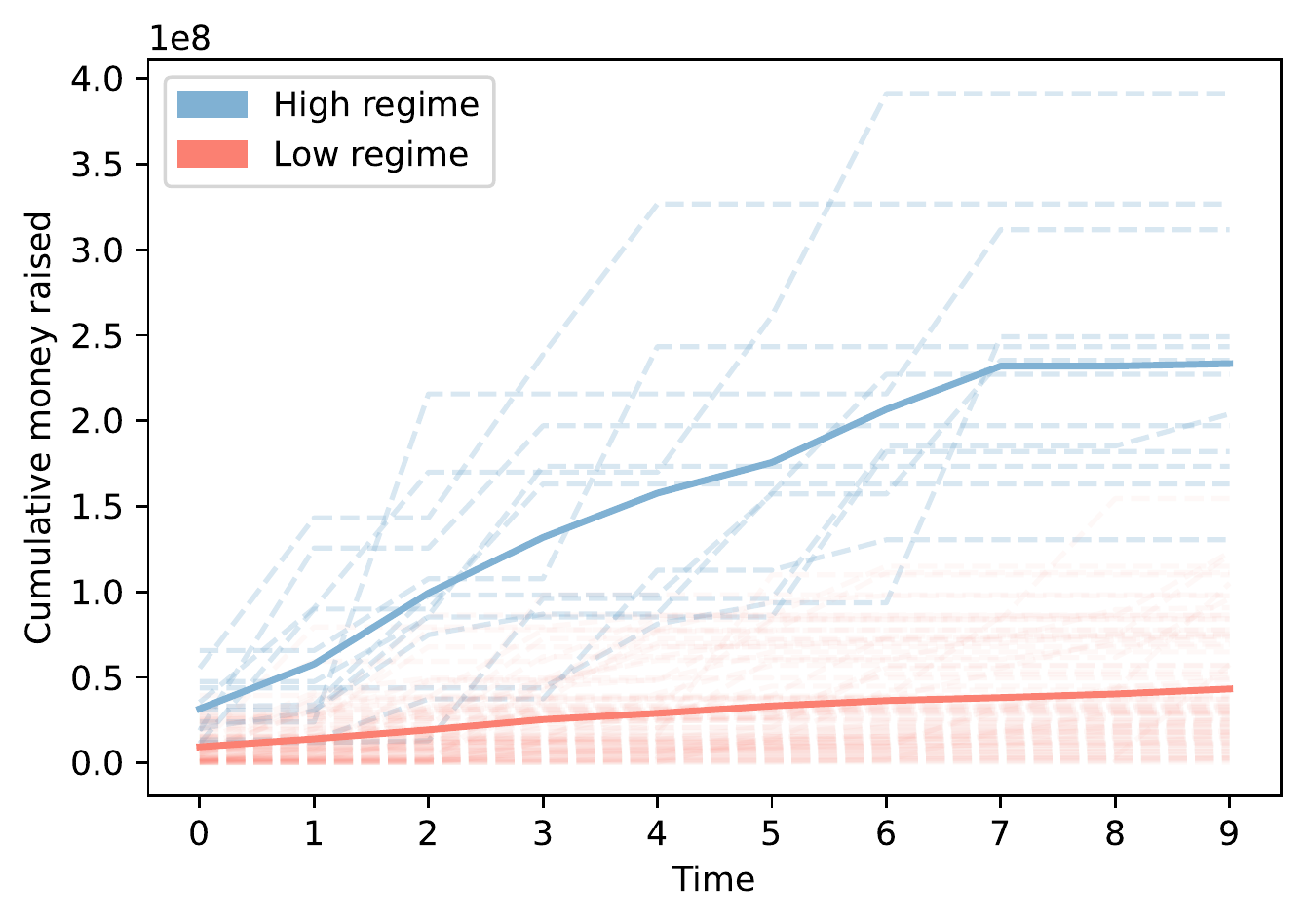}
  \caption{$k$-means functional clustering ($k=2$) of the funding trajectories of firms belonging to the drug discovery sub-sector, observed for 10 years after the first recorded round of investments. Blue and red dashed lines represent firms in the high (`successful') and low regimes, respectively. Bold curves represent cluster centroids.}
  \label{fig:funclus}
\end{figure}

Our first definition of success is based on separating firm funding trajectories in two clusters, identifying high (successful) and low investment regimes. Because of the heterogeneity among healthcare sub-sectors, we run a functional $k$-means clustering algorithm \cite{jacques2014functional} with $k=2$ separately for each sub-sector, using $100$ random initialization of the centroids. As an example, Figure \ref{fig:funclus} shows the results of the clustering algorithm for the drug discovery sub-sector. Throughout all sub-sectors, the algorithm partitions 519 (16.89\%) firms in the high-regime cluster and 2553 firms in the low-regime cluster. This binary definition of success is therefore rather conservative, as a small minority of firms are labelled as successful.

We consider a logistic regression model defined as:
\begin{equation}
\label{eq:log}
    \log\left(\frac{P(y_i=1)}{1-P(y_i=1)}\right)=\beta_0+\sum_{j=1}^{p} \beta_j x_{ij} \quad i=1,\dots n
\end{equation}
where $n$ is the number of observations; $y_i$, $i=1,\dots n$, are the binary responses indicating membership to the high ($y_i=1$) or low ($y_i=0$) regime clusters; $\beta_0$ is an intercept and $x_{ij}$, $i=1,\dots n$, $j=1,\dots,p$ ($p=7$), are scalar covariates. We fit all possible configurations of this model, comprising one covariate per group, and select the best ones in terms of their log-likelihood.

\begin{figure}
    \includegraphics[width=0.96\textwidth]{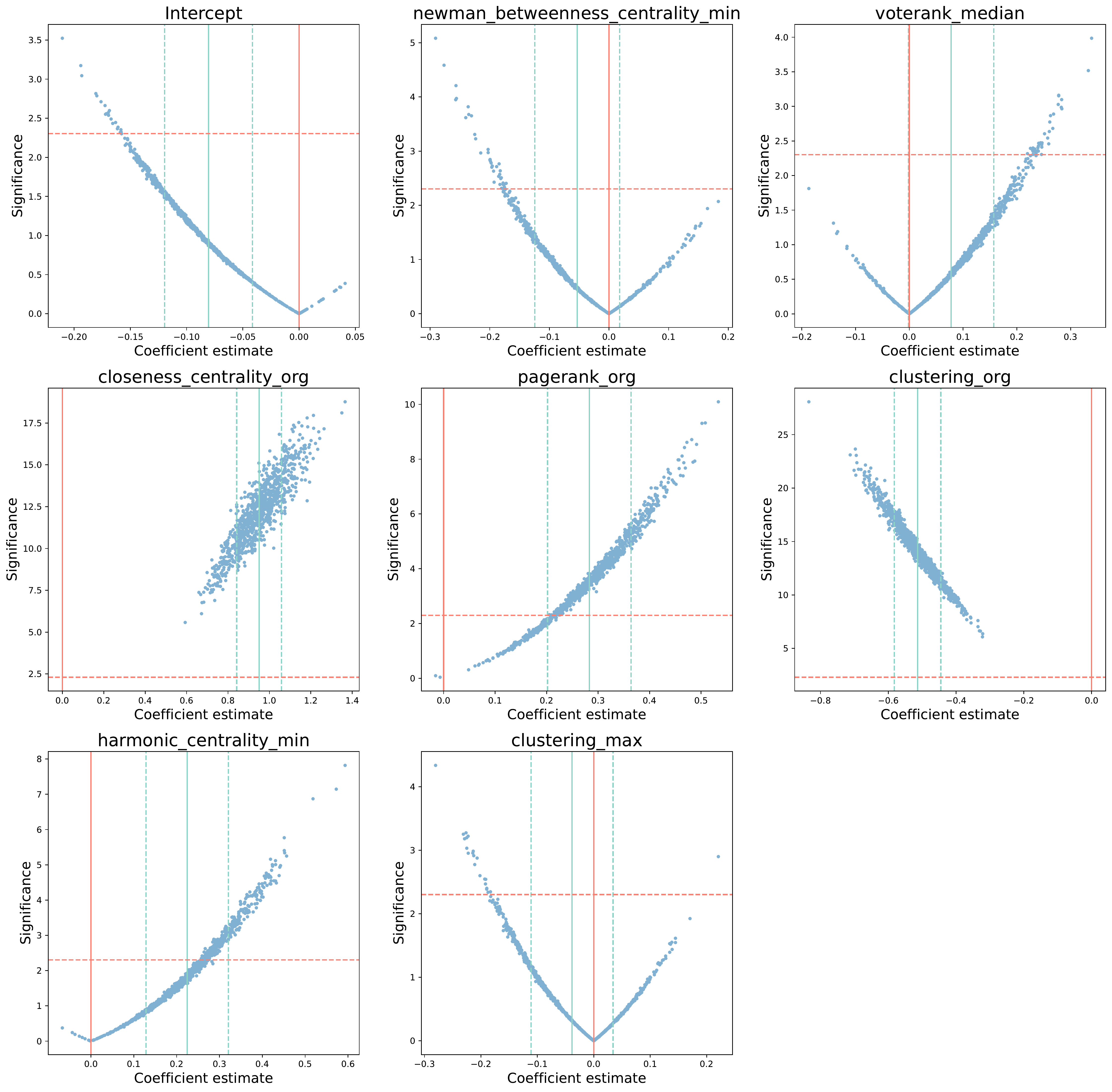}
  \caption{Scatter plots of logistic regression coefficient estimates (horizontal) and significance (vertical; $-log(p$-$value)$). The eight scatter plots correspond to the intercept (upper left) and the covariates selected in each of the seven covariate groups. Each point in a scatter plot represents one of $1000$ fits run on data balanced by subsampling the majority low-regime class. Light blue solid lines mark averages across the fits, and light blue dashed lines are $\pm1$ standard deviations about them. Red vertical solid lines mark $0$ coefficient values. Red horizontal dashed lines mark significance values associated to a p-value of 0.1.}
  \label{fig:unlogreg}
\end{figure}

Since the data set is unbalanced, results on the best model configuration, which has a log-likelihood of $-862.98$ and a McFadden's pseudo $R^2$ \cite{mcfadden1973conditional} of $0.1402$, are bound to be driven by the majority class. To mitigate this problem, for $1000$ independent times, we randomly subsample the more abundant low-regime class as to match the numerosity of the high-regime one, and re-fit the logistic regression using the best combination of regressors found on the whole data set
\cite{hastie2009elements}. The average log-likelihood across these fit replications is $-418.37$, which is substantially higher than the score of the unbalanced fit, while the average McFadden's pseudo $R^2$ is $0.1499$, with a maximum of $0.1970$. Figure \ref{fig:unlogreg} displays the scatter plots of the magnitude of the estimated coefficients and their significance. Firms' closeness (group 3) and PageRank (group 4) centrality show a positive and significant impact on the probability of belonging to the high-regime class. Interestingly, firms' clustering coefficient (group 5) seems to have a negative and significant impact. This is probably due to the conspicuous number of more peripheral firms which do not belong to the giant component of our network, and have a clustering coefficient equal to 1.  The harmonic centrality of the least important investor in a firm (group 6) seems to have a positive, mildly significant effect. Finally, the covariates selected to represent the groups 1, 2 and 7 do not show a significant association with the binary response.

Next, we investigate whether the binary definition of success derived from firms' funding trajectories is related to a more standard definition, i.e., their eventual exit in IPO, acquisition or merger. This is captured by the confusion matrix in Table \ref{tab:comparingbinary}. Notice that the overall accuracy is quite large (0.71) and the precision is 0.57. This means that, if we classified a firm as successful according to our definition, there would be a good chance that it would be selected as successful also by standard definitions. Nevertheless, the number of false negatives is not negligible (recall is only 0.31). 

\begin{table}
    \caption{Confusion matrix for standard vs trajectory-based binary definition. Accuracy: 0.71; recall: 0.31; precision: 0.57}
    \label{tab:comparingbinary}
    \centering
    \begin{tabular}{l|ll}
    \toprule
         & High-regime class & Low-regime class\\
    \midrule
        IPO/Acquired/Merged & 294 & 664\\
        No IPO/Acquired/Merged & 225 & 1889 \\
    \bottomrule
    \end{tabular}
\end{table}

\subsubsection*{Aggregate amount of money raised}
The evidence of a relationship between the success of a firm and the network features obtained using our trajectory-based binary response is promising. However, our binary definition of success is very rough and the unbalance in the data forces us to run the analysis relying on reduced sample sizes. Moreover, the confusion matrix in Table \ref{tab:comparingbinary} highlights a less-than-perfect match with a more standard binary definition, suggesting that we may be missing some important components of what makes a firm successful. Thus, we next consider a scalar proxy for success, defined as the cumulative end point of a firm's funding trajectory, i.e.~the sum of the investments it has received through a 10 year window since the first investment. 

We consider a regression model
defined as:
\begin{equation}
\label{eq:linreg}
    y_i =\beta_0+\sum_{j=1}^{p} \beta_j x_{ij} + \sum_{c=1}^{k} \gamma_c x_{ic} +\epsilon_i \quad i=1,\dots,n
\end{equation}
where $n$ is the number of observations; $y_i$, $i=1,\dots n$, are the scalar responses (aggregate amount of money raised); $\beta_0$ is an intercept; $x_{ij}$, $i=1,\dots,n$, $j=1,\dots,p$ ($p=7$), are scalar covariates; $x_{ic}$, $i=1,\dots,n$ and $c=1,\dots,k$ ($k=5$), are scalar controls and $\epsilon_i$, $i=1,\dots n$, are i.i.d.~Gaussian model errors. As for the logistic regression, we fit all possible configurations of this model, comprising one covariate per group. However, in order to properly gauge the network features, in all the fits we include controls for the size of the first investment (a scalar variable) and the industry of the firms (a categorical variable indicating the membership to a given market sector). The left column of Table \ref{tab:linreg} shows results for the fit of the best model configuration in terms of $R^2$. Additional fits with different model specifications (also controlling for firms' geographical location and year of birth) produce similar results (see Table \ref{tab:linreg_many_controls} in the supplementary material). In particular, we also run a linear regression using as response differential money raised; that is, the difference between aggregate money raised and the amount of money received in the first investment (in this specification the latter is {\em not} included as a control). Results for this fit are shown in the right column of Table \ref{tab:linreg}. As in the case of the logistic regression for our binary outcome, and for both the scalar responses considered, the covariate from group 3 (here, the maximum of the average neighborhood degree among a firm's investors) has a positive and significant effect, while the covariate from group 5 (a firm's clustering coefficient) has a negative and significant one. The covariate from group 1 (here, the median of the investors' PageRank) also has a positive and significant effect. The covariate from group 4 (firms' eigenvector centrality) has a positive and significant effect, but only with differential money raised as response. The other covariates (from groups 2, 6 and 7) do not show a significant association with the scalar responses. Notably, the regressions explain approximately $48\%$ and approximately $22\%$ of the in-sample variability of aggregate and differential money raised, respectively.

\begin{table}
\centering 
\caption{Linear regression results} 
\label{tab:linreg} 
\begin{tabular}{@{\extracolsep{5pt}}lcc} 
\toprule
 & \multicolumn{2}{c}{\textit{Response:
 }} \\ 
\cline{2-3} 
\\[-1.8ex] & Agg. money raised (log) & Diff. money raised (log) \\ 
\midrule
pagerank\_median (log) & 0.1186$^{***}$ & 0.2645$^{***}$ \\ 
                 & (0.044) & (0.063) \\ 
                 & &\\ 
voterank\_max & -0.0231 & 0.1116\\ 
              & (0.065) & (0.094)\\ 
              & &\\ 
average\_neighbor\_degree\_max (log) & 0.3084$^{***}$ & 0.9215$^{***}$\\
                               & (0.058) & (0.075)\\
                               & &\\
eigenvector\_centrality\_org & 0.0316 & 0.0721\textsuperscript{**} \\
                                & (0.021) & (0.030) \\
                               & &\\ 
clustering\_org & -0.0672$^{**}$& -0.2436$^{***}$\\
                & (0.030) & (0.041)\\
                & &\\
harmonic\_centrality\_median & 0.0584 & -0.1171\\
                               &(0.073)& (0.104)\\
                               & &\\
clustering\_min & 0.0005  & -0.0480\\
                              &        (0.035) & (0.050)\\
                             & &\\
Intercept & 16.2843$^{***}$ & 15.7837$^{***}$\\
         & (0.176) & (0.252)\\
         & & \\
\midrule
Observations & 1921 & 1917\\ 
R$^{2}$ & 0.485 & 0.217 \\ 
Adjusted R$^{2}$ & 0.482 & 0.213 \\ 
F Statistic & 149.90$^{***}$ (df = 12; 1908) & 48.11$^{***}$ (df = 11; 1905)\\ 
\midrule
\textit{Note:}  & \multicolumn{2}{r}{$^{*}$p$<$0.1; $^{**}$p$<$0.05; $^{***}$p$<$0.01} \\
\bottomrule
\end{tabular} 
\end{table} 

\begin{figure}
    \centering
    \includegraphics[width=0.6\textwidth]{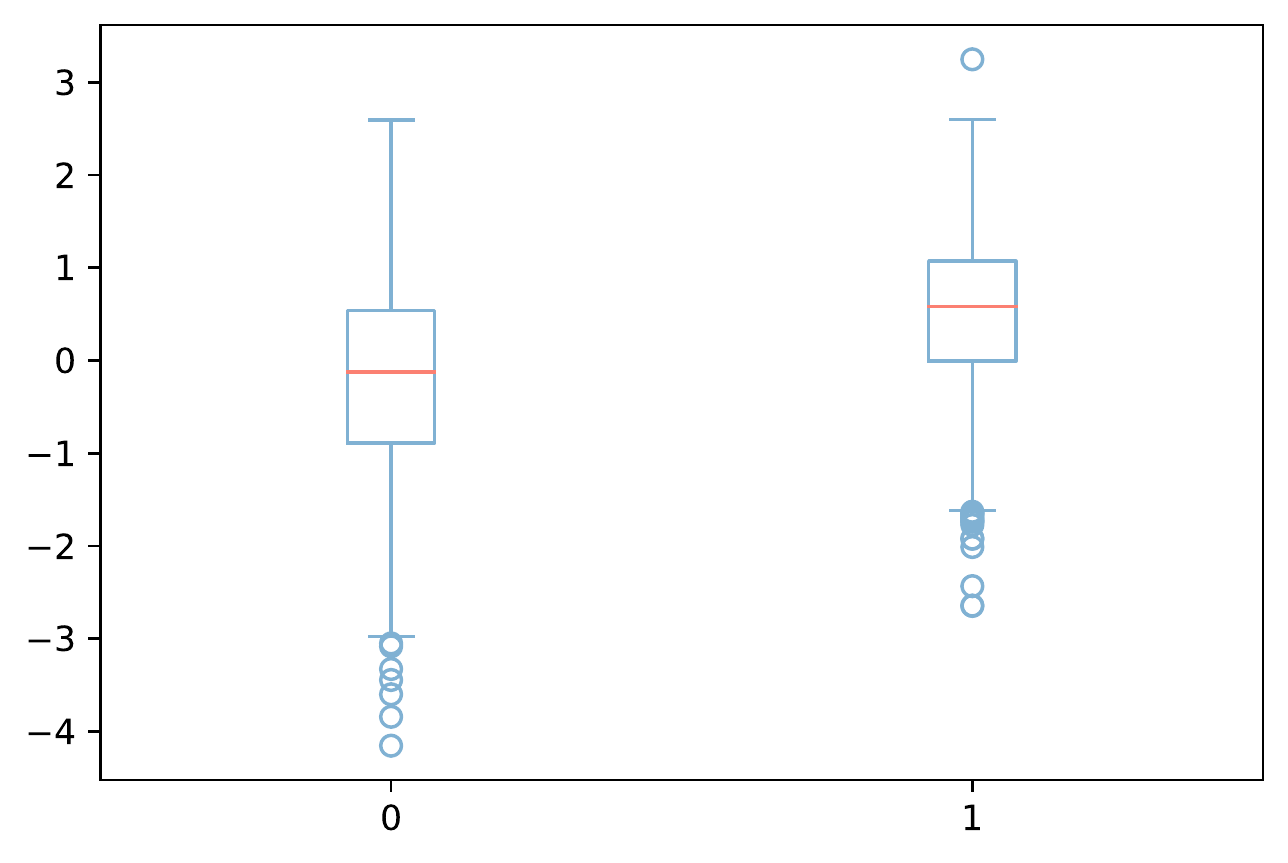}
    \caption{Boxplots of aggregate money raised (log) for `traditionally successful' (right; 1) and `traditionally unsuccessful' (left; 0) firms.}
    \label{fig:sortcum}
\end{figure}

Next, as we did for our binary response, we assess whether aggregate money raised is linked to the standard definition of success (exit in IPO, acquisition or merger within $10$ years from the first investment). Figure \ref{fig:sortcum} contrasts the boxplots of aggregate money raised between `typically defined' successful and unsuccessful firms. Firms that are successful based on the standard definition indeed do appear to raise more money (higher median, shorter left tail). This is partially expected, since the highest amount of money is usually invested in the latest stages, which are more likely to happen if the firm is promising and headed towards a successful outcome (an IPO, for instance). However, some successful exits, such as acquisitions, may also happen at an early stage of the firm life-cycle -- resulting in an exchange of shares but not in fresh new money for the firm. This may explain why our binary classification in low and high regimes misses many `typically successful' firms; the majority of the latter correspond to acquisitions, not IPOs.

\subsubsection*{Exploiting the trajectories}
Our scalar outcome, which measures the aggregate money raised by a firm within a period of $10$ years (the selected window size), does not capture how the investments in the firm distribute across such period -- something that may be very important in delineating success. Moreover, using aggregate money raised implicitly assumes that the right time to investigate the dependence of success on network features is at the end of the period considered.

We tackle these issues by refining the target outcome and considering the full cumulative investment trajectories -- instead of their end point. Thus, we run a function-on-scalar regression \cite{kokoszka2017introduction}, modeling trajectories as a function of the same set of covariates selected with our procedure for the aggregate money raised response. The function-on-scalar regression can be written as:
\begin{equation}
    \label{eq:fun_reg}
        Y_i(t) = \beta_0(t) + \sum_{j=1}^{p} \beta_j(t) x_{ij} + \epsilon_i(t) \quad i=1,\dots n
    \end{equation}
where $n$ is the number of observations; $Y_i(t)$, $i=1,\dots n$, are the aligned trajectories; $\beta_0(t)$ is a functional intercept; $x_{ij}$, $i=1,\dots n$ and $j=1,\dots,p$ ($p=7$), are scalar covariates, and $\epsilon_i(t)$, $i=1,\dots n$, are i.i.d. Gaussian model errors.

\begin{figure}
    \includegraphics[width=0.96\textwidth]{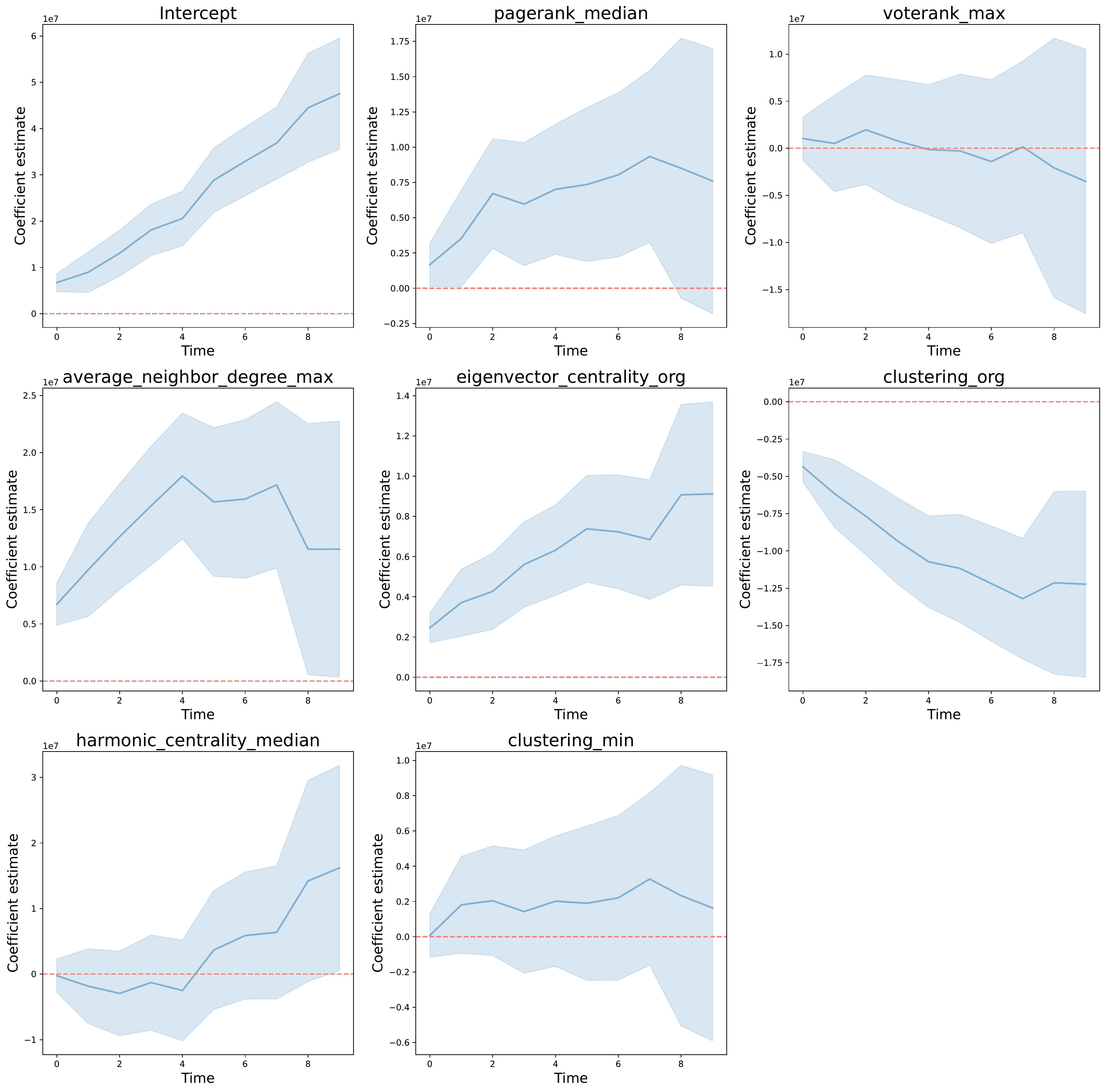}
  \caption{Function-on-scalar regression.
  Blue solid lines represent coefficient curve estimates, surrounded by blue confidence bands (constructed through point-wise 
  estimated standard errors, 
  95\% confidence level). Red dashed lines mark 0. The estimated intercept can be interpreted as the sheer effect of time.}
  \label{fig:fosreg}
\end{figure}

In this model, the regression coefficient of a scalar covariate $j$, $\beta_j(t)$, is itself a curve describing the relationship between the covariate and the functional response, which varies along its domain (time). Together with the functional coefficients, we also estimate their standard errors, which can be employed to build confidence bands around the curves~\cite{goldsmith2016refund}. Figure \ref{fig:fosreg} shows the estimated functional coefficients and their 95\% confidence bands. Results of the fit are in line with the previous ones, showing positive and significant effects for the covariates from groups 1, 3 and 4, and a negative and significant effect for the covariate from group 5. Interestingly, the positive effect of the covariate from group 1 is increasing, but its derivative is decreasing over time -- leveling off after approximately 2 years from the first investment. Also, the positive effect of the covariate from group 3 is initially increasing but, around 4 years from the first investment, starts to decrease. The positive effect of the covariate from group 4 and the negative effect of the covariate from group 5 both increase in size across the whole time domain. Finally, the effects of the covariates from groups 2, 6 and 7 are not statistically different from 0.

\subsection*{Stability checks} \label{subsec:stability}
In order to validate our previous results we extend the analysis in two ways. First, we vary the length of the funding trajectories, re-running our pipelines for all window sizes between $5$ and $12$ years. Second, we take advantage of the correlation structure characterizing the covariates at our disposal to measure the stability of coefficient estimates with respect to perturbations in model configurations.

\begin{figure}
    \centering
    \includegraphics[width=0.96\textwidth]{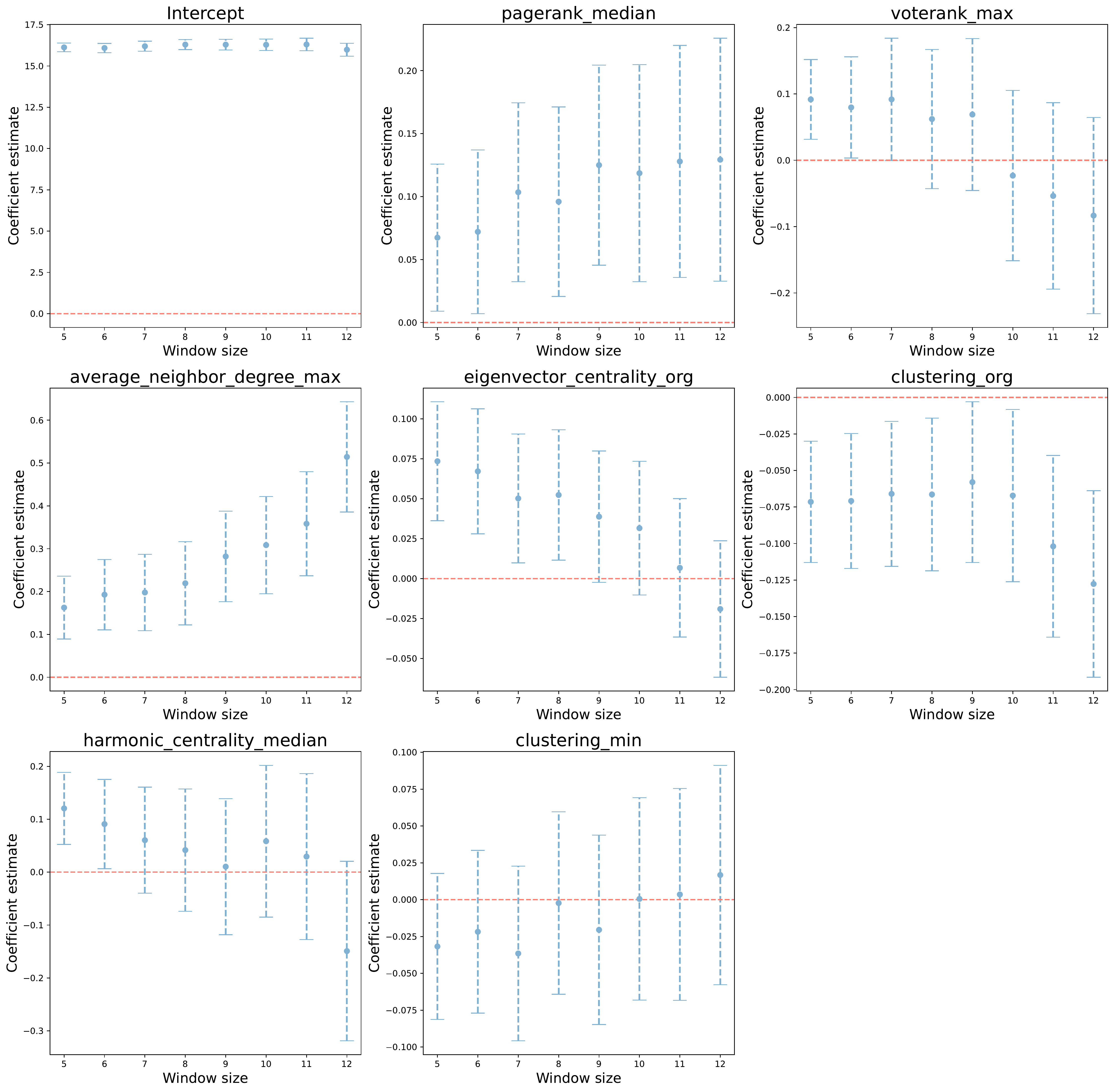}
    \caption{Stability analysis under window size changes; linear regression. Blue dots represent coefficient estimates for each window size, with $\pm 1.96\times SE$ bands (dashed blue vertical lines). Horizontal red dashed lines mark 0. The number of firms included in the fits decreases as the window size increases, possibly introducing a selection bias.}
\label{fig:stablinreg}
\end{figure}

We note that varying the window size induces a change in the number of firms in our sample (see Figure \ref{fig:n_firms}). This is due to the fact that, when aligning firms to the year of their first investment, we consider also firms which are born very recently. The number of firms that already have a trajectory long enough to cover the entire window depends on the size considered; the longer the window, the smaller the number of firms that can be used in the fits.
Figure \ref{fig:stablinreg} shows the variation in coefficient estimates when fitting the linear regression in Equation \ref{eq:linreg} with different window sizes. The coefficients for covariates from groups 1 and 3 remain distinctly positive for all window sizes. The coefficients for the covariate from group 5 remain distinctly negative for all window sizes. In contrast, the coefficients for covariates from groups 2, 4, 6 and 7 change sign depending on the window size, and have 95\% confidence intervals that overlap $0$ for most or all window sizes, indicating less stable effects. Figure \ref{fig:stablogreg} and \ref{fig:stabfosreg} show the same analysis conducted for the logistic and the function-on-scalar regressions in Equations \ref{eq:log} and \ref{eq:fun_reg}, respectively.

\begin{figure}
   \includegraphics[width=0.96\textwidth]{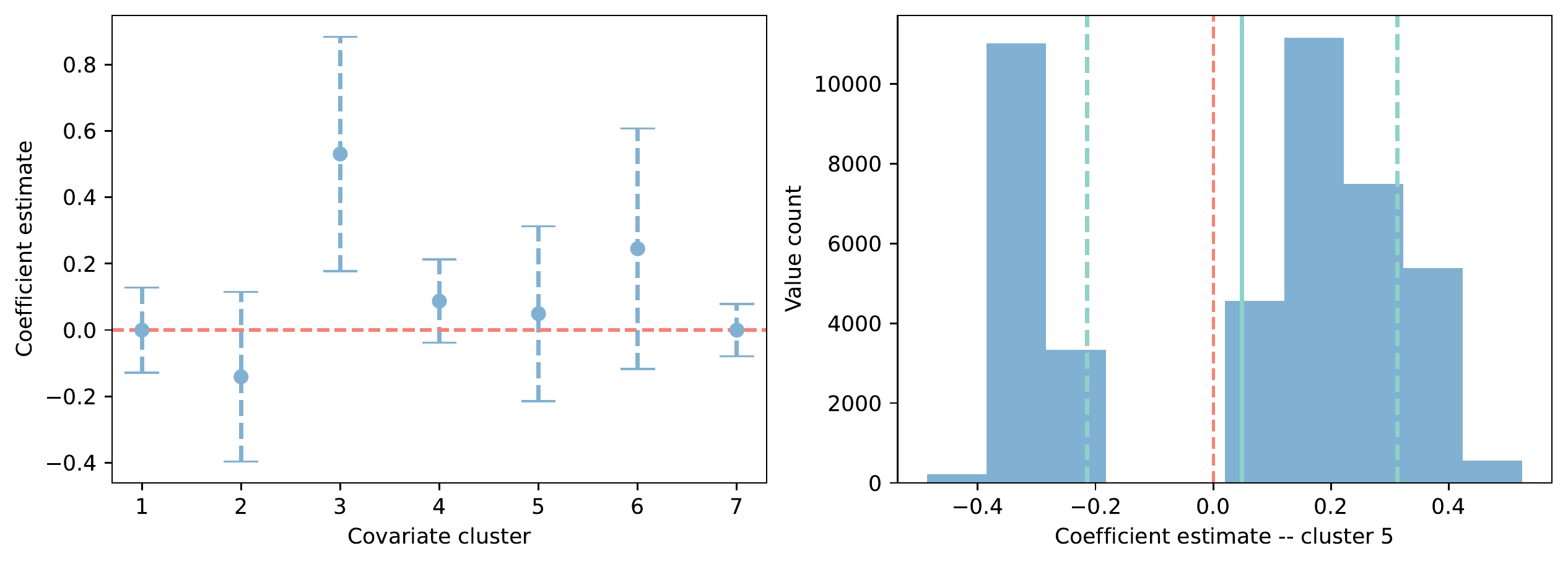}
   \caption{Stability analysis under perturbations of the model specification; linear regression.
   Left panel: blue dots represent averages of the
   coefficient estimates obtained changing the covariates considered in each group, with $\pm1$ standard deviation bands (dashed blue vertical lines). The horizontal red dashed line marks 0. 
   Right panel: distribution of the coefficients estimates associated to covariates from group 5. While $0$ (dashed vertical red line) is within $1$ SD (dashed vertical blue lines) from the average (solid vertical blue line), the distribution is bimodal with a trough about $0$.}
   \label{fig:stabpert}
\end{figure}

Figure~\ref{fig:stabpert} (left panel) shows the variation in coefficient estimates when perturbing the model configuration; that is, when taking all possible combinations of seven covariates, one from each group. In this analysis the window size is held fixed at $10$ years. Notably, for group $3$ the average coefficient estimate is more than one standard deviation above $0$, indicating that the positive effect of the group tends to persist regardless of the specific covariates selected within the groups. In contrast, the effects of the other groups appear to depend, in size and sign, on the model configuration. This must be interpreted with care though. For instance, in this analysis group 1 and group 5 covariates show both positive and negative coefficient estimates depending on the model configuration, suggesting a less consistent role for these groups than for group 3. However, $pagerank\_median$ does in fact have a significant positive effect within the context of the `best' linear regression configuration (see Table~3) -- which is also confirmed when considering the function-on-scalar regression (see Figure~\ref{fig:fosreg}). Also, and rather interestingly, the distribution of coefficient estimates for group 5 present a positive and a negative mode -- with a trough about $0$ (Figure~\ref{fig:stabpert}, right panel). This suggests that the group has indeed a consistent effect across model configurations, except that the sign will depend on which among the group members carries such effect -- because some of the highly correlated covariates in the group are in fact counter-varying.

\section*{Discussion}
By exploiting and combining techniques from the fields of network and functional data analysis, we propose progressively more refined definitions of a firm's success, and associate them with different network features through regression fits.

Logistic regression results for our binary outcome suggest a strong role for centrality measures belonging to groups 3 (firms and big investors), 4 (firms' eigenvector centrality measures), and 5 (firms as knowledge bridges) -- and a weaker role for centrality measures in group 6 (harmonic centrality measures). In terms of `best' representatives selected within such groups, a firm's closeness centrality from group 3 reflects the width of its investors' portfolio (if a firm is part of a big portfolio, its separation from other firms within the network will be lower), PageRank from group 4 is a proxy of a firm's importance and plays a role similar to eigenvector centrality in undirected networks,
and a firm's clustering coefficient from group 5 reflects tightness in community links. Concerning its estimated negative impact on success, we note that a firm's clustering coefficient is negatively correlated to its number of investors -- both because clustered firms typically belong to portfolios characterized by a high redundancy of investors, and because our network contains many isolated firms (outside the giant component) whose clustering coefficient equals 1, and who are unlikely to succeed.

Linear regression results for our scalar outcome confirm a strong role for groups 1, 3 and 5. The covariate selected from group 1 and group 4, the median of investors' PageRank and the firms' eigenvector centrality respectively, have positive estimated effects -- suggesting that influential median investors may favor a firm's success, as well as a firm's own influence. The covariate selected from group 3, the maximum of investors' average neighbor degrees, also has a positive estimated effect -- suggesting that for a firm's success forming many connections is not as critical as being connected with investors that are themselves strongly connected, as this may increase the level of capitalization in a later stage of the firm's life. The covariate selected from group 5 is again a firm's clustering coefficient, with a negative estimated effect on success. Our stability analysis also provides evidence that the effects of covariates from groups 3 and 5 are consistent across model specifications.

Function-on-scalar regression results for our functional outcome also confirm a strong role for groups 1, 3, 4 and 5. Interestingly, when profiled over time through this richer analysis, the positive effects of the median of investors' PageRank (group 1) and of the maximum of investors' average neighbor degrees (group 3) increase early in the life of a firm -- but then level off. In contrast, the positive effect of a firm's eigenvector centrality (group 4) and the negative effect of its clustering coefficient (group 5) increase throughout the temporal domain. This may mean that being connected to important and well connected investors (i.e., those with a median PageRank and large average neighbor degree) is more important early on, whereas being in a far-reaching portfolio of investors (i.e., having a small clustering coefficient and high eigenvector centrality) has a stronger impact on success later in the life of a firm. 

Our analysis can be expanded in several ways. First, we limit our study to the healthcare sector, while it may be interesting to investigate other market sectors, and compare the results. Second, meso-scale communities may be analyzed in terms of their longitudinal evolution, as to characterize successful clusters of firms from a topological point of view. Third, in our analysis we gather information from the first round of funding and predict the future success of the firm, but it may be interesting to do so experimenting with different funding rounds, or with models that capture the dynamic evolution of the network (e.g. topological data analysis \cite{hensel2021survey}). This would also shed light on whether the way investment decisions travel within the network of investors has an impact on future success of firms (an application of graph kernels \cite{rosenfeld2020kernel}). Also, we could consider the network as being formed by the decision of rational agents operating within an economic framework (e.g., reinforcement learning \cite{jiang2018graph}) and employ concepts taken from the field of game theory and goal recognition design to study investment decisions among partially informed agents \cite{keren2020information}. 


\begin{backmatter}

\section*{Acknowledgements}
We are grateful to Enrico Stivella and Antonio Ughi for useful feedback.

\section*{Funding}
F.C., C.E., G.F., A.M. and L.T.~acknowledge support from the Sant'Anna School of Advanced Studies. 
F.C.~acknowledges support from Penn State University. 
G.R.~acknowledges support from the scheme \enquote{INFRAIA-01-2018-2019: Research and Innovation action", Grant Agreement n. 871042, "SoBigData++: European Integrated Infrastructure for Social Mining and Big Data Analytics}.

\section*{Availability of data and materials}
The data that support the findings of this study are available from CB Insights but restrictions apply to the availability of these data, which were used under license for the current study, and so are not publicly available. Data are however available from the authors upon reasonable request and with permission of CB Insights. Code for replication of our study is shared in the GitHub repository \url{https://github.com/testalorenzo/crunch_net}


\section*{Competing interests}
The authors declare that they have no competing interests.


\section*{Authors' contributions}
All authors conceived ideas and analysis approaches. C.E., M.G.~and L.T.~retrieved and processed data, implemented pipelines and performed statistical analyses. All authors interpreted findings. F.C., M.G.~and L.T. wrote the manuscript. F.C., F.G., A.M.~and G.R.~supervised the research.



\bibliographystyle{bmc-mathphys} 
\bibliography{bmc_article}      









\section*{Additional Files}
  \subsection*{Supplementary materials are available}


\end{backmatter}
\end{document}